\documentclass[aps,prl,twocolumn,showpacs,superscriptaddress,nofootinbib,10pt]{revtex4-1}

\usepackage{acronym}
\usepackage{amsfonts}
\usepackage{amsmath}
\usepackage{amssymb}
\usepackage{bm}
\usepackage{color}
\usepackage{epsfig}
\usepackage{etoolbox}
\usepackage{float}
\usepackage{graphicx}
\usepackage{hyperref}
\usepackage{indentfirst}
\usepackage{latexsym}
\usepackage{mathrsfs}
\usepackage{multirow}
\usepackage{rotating}
\usepackage{subfigure}
\usepackage{tabls}
\usepackage{verbatim}

\newcommand{\mb}[1]{\mbox{\boldmath $#1$}}

\newcommand{\salto}[1]{\left[\,#1\,\right]^{}_{p}}

\def \met  {\mbox{g}}

\def \rtor {{r^{}_{\!\ast}}}

\newcommand{\BH}{{\mbox{\tiny BH}}}
\newcommand{\GAUGE}{{\mbox{\tiny G}}}

\newcommand{\LO}{{\mbox{\tiny L}}}
\newcommand{\REG}{{\mbox{\tiny R}}}
\newcommand{\RW}{{\mbox{\tiny RW}}}
\newcommand{\ZM}{{\mbox{\tiny ZM}}}

\definecolor{orange}{rgb}{1,0.5,0}

\newcommand{\cm}[1]{{\color{red}~\textsf{CITEME!}}}

\begin{document}

\title[Overcoming the Gauge Problem for the Gravitational Self-Force]{Overcoming the Gauge Problem for the Gravitational Self-Force}

\author{Priscilla Canizares}
\affiliation{Institute of Astronomy, Madingley Road, Cambridge, CB3 0HA, United Kingdom}

\author{Carlos F. Sopuerta}
\affiliation{Institut de Ci\`encies de l'Espai (CSIC-IEEC), Campus UAB, Facultat de Ci\`encies, 08193 Bellaterra, Spain}

\date{\today}

\begin{abstract}
The gravitational waves emitted by binary systems with extreme-mass ratios carry unique astrophysical information that can only be detected by space-based detectors like eLISA.  To that end, a very accurate modelling of the system is required. The gravitational self-force program, which has been fully developed in the Lorenz gauge, is the best approach we have so far. However, the computations required would be done more efficiently if we could work in other gauges, like  the Regge-Wheeler (RW) one in the case of Schwarzschild black holes. In this letter we present a new scheme, based on the Particle-without-Particle formulation of the field equations, where the gravitational self-force can be obtained from just solving individual wave-type equations like the master equations of the RW gauge. This approach can help to tackle the yet unsolved Kerr case.
\end{abstract}

\pacs{04.30.Db, 04.40.Dg, 95.30.Sf, 97.10.Sj}

\keywords{EMRIs, self-force, gravitational radiation}

\maketitle

-\textit{Motivation.} Extreme-Mass-Ratio Inspirals (EMRIs) are one of the main sources of gravitational waves (GWs) for space-based detectors like the eLISA concept proposed in {\em The Gravitational Universe}~\cite{AmaroSeoane:2013qna} -- the science theme selected by the European Space Agency for its future L3 mission. EMRIs  are binary system which consist of a stellar compact object (SCO; with a mass range $m_{\ast}\sim 1-50M_{\odot}$) orbiting a massive black-hole (MBH; with a mass range $M_{\bullet}\sim 10^{5-7}M_{\odot}$).  In the regime where the dynamics is driven by GW emission, the SCO inspirals into the MBH sweeping through the eLISA frequency band, and mapping the MBH spacetime onto the structure of the GWs in great detail.  EMRIs GW signals are also very long, since they emit around $\sim 10^{5}$ GW cycles during  the last year before plunge. Hence, GW observations are a powerful tool for astrophysics, cosmology, and fundamental physics~\cite{AmaroSeoane:2007aw,*Gair:2008bx,*AmaroSeoane:2010zy}.  However, due to the complexity of EMRI GW signals, we need precise theoretical waveform templates -- accurate enough to be in phase with the emitted GWs within a detector frequency bin, to extract the physical parameters of the system from the detector's data stream.

-\textit{EMRI modelling.} The gravitational {\em self-force} (GSF) program is the most accurate approach proposed to model EMRIs~\cite{Barack:2009ux,*Poisson:2011nh}.  The basic framework is provided by BH perturbation theory, where the spacetime metric is the one of the MBH spacetime (with metric $\met^{\BH}_{\mu\nu}$, $\mu,\nu,\ldots=0-3$) plus perturbations,  $h^{}_{\mu\nu}\ll \met^{\BH}_{\mu\nu}$, generated by the SCO.  The main challenge is to deal with the backreaction of the SCO perturbations on the SCO trajectory itself.  The backreaction is described by a local force, the {\em self-force}, acting on the SCO and deviating it from the otherwise geodesic motion in the BH background.  The main technical difficulties are related to the fact that SCO is described as a point-like object, and hence the generated metric perturbations turn out to be singular at the SCO trajectory. The first consistent formulation of the GSF was done by Mino, Sasaki and Tanaka~\cite{Mino:1997nk} and Quinn and Wald~\cite{Quinn:1997am}, who derived a formal expression of the GSF and of the equation of motion (the \emph{MiSaTaQuWa} equation) in the Lorenz (L) gauge,  $\met_{\BH}^{\rho\sigma}h_{\mu \nu;\sigma} =0$, at linear order in the mass ratio $\mu=m_{\ast}/M_{\bullet}$ (see also~\cite{Detweiler:2002mi,*Gralla:2008fg}).

The computation of the GSF is a technically complex problem due to the fact that the full retarded metric perturbations $h^{}_{\mu\nu}$ diverges on the particle's worldline. Hence a regularization scheme has to be applied to remove the singular piece and compute the SCO motion.  A popular scheme was introduced in the L gauge~\cite{Barack:2002bt}, namely the {\em mode sum} regularization scheme, where one subtracts, multipole by multipole, that singular contribution given in terms of a series of regularization parameters -- known analytically both in the case of Schwarzschild and Kerr MBHs.  This program to obtain the GSF in the L gauge has been pursued in the last years with the help of different numerical techniques, in both the time and frequency domains, and the first-order GSF has been obtained in Schwarzschild~\cite{Barack:2009ey,*Barack:2010tm,*Barack:2008ms,*Akcay:2010dx,*Akcay:2013wfa} (for generic orbits) and Kerr~\cite{Shah:2010bi} (circular case only). However,  GSF computations for generic orbits in Kerr, and specially for the computation of the second-order GSF -- which we may need to know for an adequate exploitation of EMRI GW observations~\cite{Flanagan:2010cd}, still need further developments.

Although the L gauge allows for a consistent definition of the GSF, the perturbative Einstein field equations are in general coupled -- in contrast with, for instance, the Regge-Wheeler (RW) gauge~\cite{Regge:1957rw}.  The difference in the choice of gauge can be even more important in  Kerr, where it would be very convenient to perform the computation in gauges like the {\em radiation gauge}. In this case the metric perturbations can be obtained through a reconstruction procedure~\cite{Chrzanowski:1975wv,*Kegeles:1979an,*Wald:1978vm,*Stewart:1978tm}, applied to the solution of completely decoupled master equations of the Teukolsky type~\cite{Teukolsky:1972le,*Teukolsky:1973ap}. In this  regard, Barack and Ori~\cite{Barack:2001ph} argued that, under certain differentiability conditions on the vector field generating the gauge transformation, one could use the mode sum scheme~\cite{Barack:2002bt}, to regularize the full force with the same regularization parameters as in the L gauge.  Recently more studies have addressed this question~\cite{Pound:2013faa,*Gralla:2011ke,*Gralla:2011zr}.  One of the conclusions that emerges is that the RW gauge does not seem to be sufficiently smooth to allow for transformations from the L gauge, making difficult to perform computations of the GSF (see also~\cite{Hopper:2012ty}).

In this letter we present a new scheme to compute the GSF for the case of a Schwarzschild MBH.  Its main advantage is that most computations are done in the RW gauge, where we only have to solve individual wave-type equations.  The key point is to control the singularities of the RW gauge using the Particle without Particle (PwP) technique introduced in~\cite{Canizares:2008dp,*Canizares:2009ay,*Canizares:2010yx,*Canizares:2011kw,*Jaramillo:2011gu} for computations of the scalar self-force on a charged particle orbiting a Schwarzschild MBH.  In this way, we can construct the full retarded metric perturbations in the RW gauge.  Then, we discuss how we can transform to the L gauge, again by solving individual wave-type equations, where we can obtain the GSF.  Finally, we  discuss how this can help approaching the problem of the GSF in Kerr.

-\textit{Metric perturbations in Schwarzschild}. The metric perturbations $h^{}_{\mu\nu}$ in arbitrary gauge satisfy the linearized Einstein equations ($G^{}_{\mu \nu}= \delta G^{}_{\mu \nu} = 8\pi T^{}_{\mu\nu}$):
\begin{eqnarray}
-16\pi T^{}_{\mu\nu}&=&\Box^{}_{4}\psi^{}_{\mu\nu} + \met^{\BH}_{\mu\nu}\,\psi^{\rho\sigma}{}^{}_{;\rho\sigma} -2 \psi^{}_{\rho(\mu}{}^{;\rho}{}^{}_{;\nu)} \nonumber\\
&+& 2 {R}^{\BH}_{\rho\mu\sigma\nu}\psi^{}_{\rho\sigma} - 2 {R}^{\BH}_{\rho(\mu}\psi_{\nu)}{}^{\rho} \,.
\label{eq:2.2}
\end{eqnarray}
where $T^{\mu\nu}$ is the energy-momentum tensor generating the perturbations, $\psi^{}_{\mu \nu} = h^{}_{\mu \nu} - (1/2)\met^{\BH}_{\mu\nu}\,\met_{\BH}^{\rho\sigma} h^{}_{\rho\sigma}$ are the trace-reversed metric perturbations, a semi-colon denotes covariant differentiation with respect to the MBH metric $g^{\BH}_{\mu\nu}$, $\Box^{}_{4}\psi=\met^{\mu\nu}_{\BH}\psi^{}_{;\mu;\nu}$, and $R^{\mu}_{\BH\,\nu\rho\sigma}$ and $R^{\BH}_{\mu\nu}$ are the Riemann and Ricci tensor of the Schwarzschild metric respectively.  In the case of EMRIs the energy-momentum tensor accounts for the SCO and hence it is a distributional tensor with support only on the SCO trajectory ($x^{\mu} = z^{\mu}(\tau)$, being $\tau$ the SCO proper time):
\begin{eqnarray}
T^{\mu\nu} = m_{\ast}\int \frac{d\tau}{\sqrt{-\met_{\BH}}} u^{\mu}u^{\nu}\delta^4\left(x^{\beta}-z^{\beta}(\tau)\right) \,,
\label{eq:2.1}
\end{eqnarray}
where $u^{\mu}= dz^{\mu}/d\tau$ is the SCO velocity ($u^{\mu}u^{}_{\mu}=-1$) and $\met_{\BH}=\det(\met^{\BH}_{\mu \nu})$.  In the case of Schwarzschild, the metric is the warped product of a Lorentzian two-dimensional metric describing the time-radial sector, $g_{ab}$ ($a,b,\ldots=0,1$), and the metric of the two-sphere, $\Omega_{AB}$ ($A,B,\ldots=2,3$), being the warp factor $r^{2}$, where $r$ is the areal radial coordinate.  Then, using coordinates $(x^{a},\theta^{A})$, the Schwarschild metric is given by: $ds^{2}=g_{ab}dx^{a}dx^{b}+r^{2}\Omega_{AB}d\theta^{A}d\theta^{B}$. In Schwarzschild coordinates, $g_{ab}dx^{a}dx^{b} = -fdt^{2}+f^{-1}dr^{2}\,$, and $\Omega_{AB}d\theta^{A}d\theta^{B} = d\theta^{2}+\sin^{2}\theta d\varphi^{2}\,$.  The spherical symmetry is particularly useful as we can expand any tensorial quantity in (tensor) spherical harmonics~\cite{Gerlach:1979rw,*Gerlach:1980tx}.  In this letter we use scalar ($Y^{\ell m}$), vector (polar, $Y^{\ell m}_{A}$, and axial, $S^{\ell m}_{A}$), and 2-rank tensor (polar, $Y^{\ell m}_{AB}$ and $Z^{\ell m}_{AB}$, and axial, $S^{\ell m}_{AB}$) harmonics (see~\cite{Sopuerta:2005gz} for the definitions).  Then, the harmonic modes of the metric perturbations in a gauge G (in this letter G=RW, L) can be written as (we drop the harmonic indices $(\ell,m)$):
\begin{eqnarray}
&& h^{\GAUGE}_{ab} = p^{\GAUGE}_{ab}\,Y\,,\quad h^{\GAUGE}_{aA} = q^{\GAUGE}_{a}\,Y^{}_{A} + h^{\GAUGE}_{a}\, S^{}_{A}\,,\nonumber \\
&& h^{\GAUGE}_{AB} = r^2\,(K^{\GAUGE}\, Y^{}_{AB} + G^{\GAUGE}\, Z^{}_{AB}) + h^{\GAUGE}_{2}\, S^{}_{AB}\,,  
\label{eq:2.3}
\end{eqnarray}
where ($p^{\GAUGE}_{ab}\,$, $q^{\GAUGE}_{a}\,$, $h^{\GAUGE}_{a}\,$, $K^{\GAUGE}\,$, $G^{\GAUGE}\,$, $h^{\GAUGE}_{2}$) are functions of $x^{a}$ only. In the same way, we can decompose the energy-momentum tensor in harmonics and we denote the equivalent harmonic components by ($8\pi Q_{ab}\,$, $4\pi Q_a\,$, $4\pi P_a\,$, $4\pi r^2 Q^Y\,$, $4\pi Q^Z\,$, $4\pi P$).  From Eq.~(\ref{eq:2.1}), these quantities are proportional to $\delta(r-r_{p}(t))\equiv \delta_{p}$, where $r_{p}(t)$ is the radial trajectory. Since Eqs.~(\ref{eq:2.2}) are linear, the equations for the different harmonics modes are decoupled.

The RW gauge~\cite{Regge:1957rw} is characterized by imposing the following algebraic conditions: $q_{a}^{\RW} = G^{\RW} =h^{\RW}_{2}=0$.  In this gauge we can decouple Eqs.~(\ref{eq:2.2}), harmonic by harmonic, by introducing two master functions, one for each parity sector: $\Psi^{\RW}_{\ell m}$ for the axial sector and $\Psi^{\ZM}_{\ell m}$ for the polar one~\cite{Regge:1957rw,Zerilli:1970fj}.  These master functions satisty wave-type equations of the form:
\begin{equation}
\left(\Box^{}_{2}-V^{\RW/\ZM}_{\ell}(r)\right)\Psi^{\RW/\ZM}_{\ell m} = S^{\RW/\ZM}_{\ell m}\,,
\label{eq:3.1}
\end{equation}
where $\Box^{}_{2}= -\partial^{2}_{t} + f\partial^{}_{r}\left(f\partial^{}_{r}\right) = -\partial^{2}_{t}+\partial^{2}_{r_{\ast}}$ and $r^{\ast} = r+2M_{\bullet}\ln{\left(r/2M_{\bullet}-1\right)}$ is the so-called \emph{tortoise} coordinate.  The potentials $V^{\RW/\ZM}_{\ell}$ only depend on the harmonic number $\ell$ and their form can be found in~\cite{Regge:1957rw,Zerilli:1970fj}.  The source terms $S^{\RW/\ZM}_{\ell m}$ generated by the SCO have the following structure: $S^{\RW/\ZM}_{\ell m} = {\cal G}_{\ell m}^{\RW/\ZM}\delta_p + {\cal F}_{\ell m}^{\RW/\ZM}\delta'_p\,$. Consequently the master functions will have discontinuities across the SCO trajectory. After the master functions are found we can reconstruct from them the rest of metric perturbations (see, e.g.~\cite{Martel:2003jj}).

In contrast, the L gauge class is defined by the differential conditions $\met^{\rho\sigma}_{\BH}\psi^{}_{\mu\rho;\sigma} =0\,$ so the gauge is not completely fixed.  Then, Eqs.~(\ref{eq:2.2}) simplify to: $\Box^{}_{4}\psi^{}_{\mu \nu} +2 {R}^{\BH}_{\rho\mu\sigma\nu}\psi_{\rho\sigma} = -16\pi T^{}_{\mu \nu}\,$.  Here, although axial and polar modes decouple, the equations for the metric perturbations of each harmonic within each parity type, $\mb{U}$, are coupled and have the following structure: $\Box^{}_{2} \mb{U} +\Gamma\cdot \partial^{}_t\mb{U} + \Lambda\cdot\partial^{}_\rtor\mb{U}+ \Pi\cdot\mb{U} = \mb{H}\,\delta_p\,,$ where $\Gamma$, $\Lambda$, and $\Pi$ are matrices and $\mb{H}$ is a vector.  Since the source term does not contain derivatives of $\delta_{p}$ the solutions of these equations are continuous across the SCO trajectory.

\begin{figure}[ht]
\begin{center}
\includegraphics[scale=0.26]{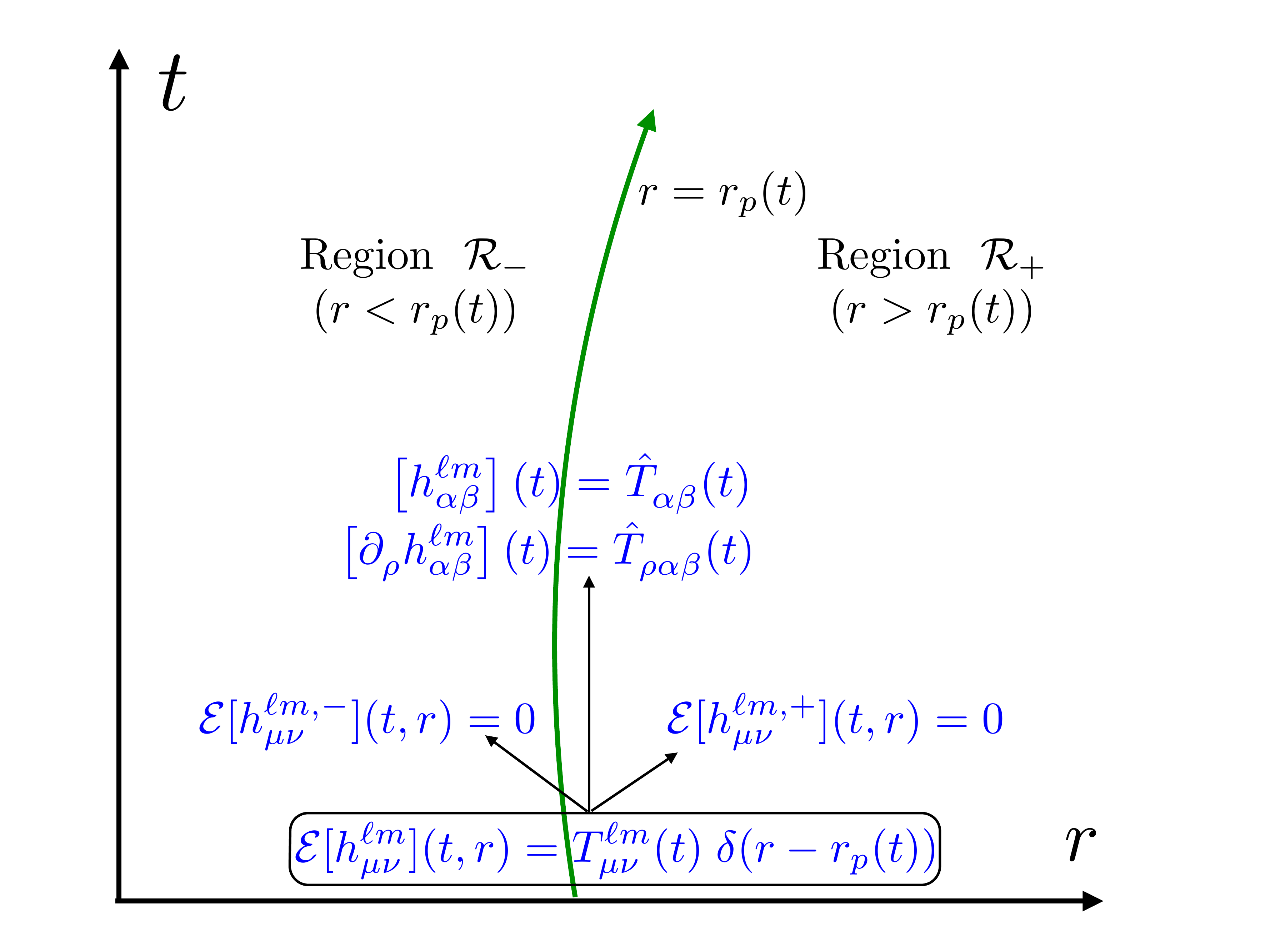}
\caption{Schematic representation of the PwP formulation. The field equations  with singular source terms (box) become homogeneous equations at each side of the particle worldline together with a set of jump conditions to communicate their solutions.}\label{FIG1}
\end{center}
\end{figure}

-\textit{The Particle without Particle Formulation}.  In general,  the full retarded metric perturbations have to be found numerically and hence, it is very convenient to formulate their equations so that we obtain smooth solutions.  However, the presence of singularities in Eqs.~(\ref{eq:2.2}), represented by Dirac delta distributions, makes the task difficult.  To overcome these problems the PwP was introduced~\cite{Canizares:2008dp,*Canizares:2009ay,*Canizares:2010yx,*Canizares:2011kw}.  It is based on a simple idea, to split the computational domain (in the $x^{a}=(t,r)$ space) into two disjoin regions (see Fig.~\ref{FIG1}): Region ${\cal R}_{-}$ to the left of the SCO trajectory ($r<r_p(t)$) and region ${\cal R}_{+}$ to the right ($r>r_p(t)$). Then, any quantity ${\cal Q}(t,r)$ that is continuous across the SCO trajectory, like the metric perturbations in the L gauge, admits a decomposition 
\begin{equation}
{\cal Q} = {\cal Q}^{-}\,\Theta^{-}_{p} + {\cal Q}^{+}\,\Theta^{+}_{p}\,,\label{PwPsplit}
\end{equation}
where we have defined $\Theta^{-}_{p} \equiv \Theta(r_p-r)$ and $\Theta^{+}_{p}\equiv \Theta(r-r_p)$, and $\Theta$ is the Heaviside step function.  Quantities that are not continuous will have jumps across the SCO trajectory.  The jump in a quantity ${\cal Q}$ is a time-only dependent quantity defined as: $[{\cal Q}](t) = \lim_{r\rightarrow r^{}_{p}(t)} {\cal Q}^{+}(t,r) - {\cal Q}^{-}(t,r) \equiv \salto{{\cal Q}}\,$.  When we apply the PwP formulation to the perturbative Einstein equations~(\ref{eq:2.2})  they transform into two sets of homogeneous equations (no matter source terms) at each region $(\pm)$, plus a set of jump conditions on the metric perturbations and their derivatives.  That is, at each region we have equations without the singular terms induced by the SCO.  Then, in the case that these equations are strongly hyperbolic -- as it happens with the L and RW gauges, we obtain smooth solutions.  Finally, the SCO appears in the communication between the two regions by enforcing the jump conditions. The spherical symmetry of the MBH background leads to jumps only in the time and radial derivatives of the metric perturbations that are not independent. In particular, for first order derivatives we find: $\salto{\partial_{t}Q^{\ell m}} = d \salto{Q^{\ell m}}/dt - \dot{r}_{p} \salto{\partial_{r}Q^{\ell m}}$. And the same happens for derivatives of higher order.  

Within the PwP formulation we analyze the perturbative equations in the L and RW gauges, together with their gauge conditions and the gauge transformation equations.  In relativistic perturbation theory, the transformation between gauges is described by a vector field $\xi^{\mu}$ in such a way that the metric perturbation associated with these two gauges are related by:
\begin{equation}
h^{\LO}_{\mu\nu}  = h^{\RW}_{\mu\nu} + 2\,\xi^{}_{(\mu;\nu)}\,. 
\label{gaugetransformationequation}
\end{equation}
The gauge vector $\xi^{\mu}$ can  be expanded in (scalar and vector) spherical harmonics as: $\xi^{\ell m}_{\mu}= (\xi_a^{\ell m}\,Y^{\ell m},\, \xi^{\ell m}_{\cal P}\,Y_{A}^{\ell m} + \xi^{\ell m}_{\cal A}\, S_{A}^{\ell m})\,$, where $\xi_a^{\ell m}$ and $\xi_{\cal P}^{\ell m}$ are the polar components and $\xi_{\cal A}^{\ell m}$ is the only axial component.

Introducing the PwP representation [Eq.~(\ref{PwPsplit})] of the L gauge metric perturbations into Eqs~(\ref{eq:2.2}) and into the L gauge conditions, we find all the jumps in the metric perturbations and their radial derivatives. As expected, the metric perturbations are continuous, i.e. $\salto{h^{\rm L}_{\mu\nu}}=0$.  The jumps in the radial derivatives, $\salto{\partial^{}_{r}h^{\rm L}_{\mu\nu}}$, which in general do not vanish, are given in terms of the harmonic components of the energy-momentum tensor, $r_{p}$ and $\dot{r}_{p}$ -- higher derivatives can be written in terms of $(r_{p},\dot{r}_{p})$ using the geodesic equations.
On the other hand, introducing the PwP representation [Eq.~(\ref{PwPsplit})] of $\xi^{\mu}$ into the gauge transformation equations~(\ref{gaugetransformationequation}), Dirac delta distributions will appear from the derivatives of $\xi^{\mu}$.  Given that the L gauge metric perturbations are regular, this implies that the RW gauge metric perturbations do not admit a presentation like in Eq.~(\ref{PwPsplit}) but they must have the following form
\begin{eqnarray}
h^{\RW}_{\mu\nu} = h^{\RW,-}_{\mu\nu}\Theta^{-}_{p} + h^{\RW,+}_{\mu\nu}\Theta^{+}_{p}+ {\cal D}^{\RW}_{\mu\nu}\delta_p\,,
\label{eq:6.3}
\end{eqnarray}
where the coefficients ${\cal D}^{\RW}_{\mu\nu}$ can be assumed to be functions of $t$ only. By looking into the perturbative Einstein equations in the RW gauge it turns out that this singular structure of the equations is needed. Actually, a key equation is the one corresponding to the trace-free polar tensor harmonic component, which has the following structure
\begin{eqnarray}
Q^{Z}\delta^{}_{p} &=& -g^{ab}p^{\RW}_{ab}  \label{eq:6.4} \\
&=& -g^{ab}p^{\RW,+}_{ab}\Theta^{+}_{p}-g^{ab}p^{\RW,-}_{ab}\Theta^{-}_{p} -g^{ab}{\cal D}^{\RW}_{ab}\delta^{}_{p}\,,\nonumber
\end{eqnarray}
where $Q^{Z} \propto m_{\ast} (f^{}_{p}/r^{2}_{p})(L^{2}_{p}/E^{}_{p})u^{A}u^{B}\bar{Y}_{AB}(\theta^{}_{p}(t),\varphi^{}_{p}(t))$, and $E_p$ and $L_{p}$ are the SCO energy and angular momentum respectively, and a bar denotes complex conjugation. This singular term vanishes identically in the case of radial trajectories ($u^A=0\Rightarrow Q^{Z}=0$), which explains why computations of the self-force in the RW gauge were only developed for radial trajectories~\cite{Barack:2002ku}.  In the case of generic trajectories, Eq.~(\ref{eq:6.4}) implies: $Q^{Z}=-g^{ab}{\cal D}^{\RW}_{ab}$ and $g^{ab}p^{\RW,\pm}_{ab}=0$, that is, the singular terms arising from the gauge transformation account for the singular term of the energy-momentum tensor.  All this happens because the right-hand side, corresponding to the perturbed Einstein tensor in the RW gauge, does not contain any derivatives of the metric perturbations, as a consequence of the RW gauge.  Therefore, unless the metric perturbations are singular, i.e. contain Dirac delta terms, there is no way to compensate the singular behaviour of the energy-momentum tensor -- this explains why we need the singular structure of $h^{\RW}_{\mu\nu}$.  By looking at the gauge transformation equation~(\ref{gaugetransformationequation}) and the rest of the perturbative Einstein equations, we can find that the relation between the coefficients ${\cal D}^{\RW}_{\mu\nu}$ and the jumps in the gauge vector field $\xi^{\mu}$ is: ${\cal D}^{\RW}_{ab} = -2\Gamma^{-1}\salto{\xi^{}_{r}}n^{}_{a} n^{}_{b}$, where $n^{}_{a}=\Gamma(-\dot{r}_{p},1)$ is a unit vector perpendicular to the SCO trajectory and $\Gamma^{-2}= f^{}_{p}(1-\dot{r}^{2}_{p}/f^{2}_{p})$ (with $f^{}_{p} = 1-2M_{\bullet}/r^{}_{p}$) is the normalization factor.  Then, thanks to the PwP formulation we control these terms and understand completely how they arise in the gauge transformation.  The gauge transformation must  be generated by a vector field $\xi^{\mu}$ that is not continuous at the SCO trajectory, but otherwise it is finite there, and then satisfying  the criteria of Gralla and Wald~\cite{Gralla:2008fg} for allowed gauge transformations.

In summary, in the PwP formulation we have the well-known vacuum Einstein perturbative equations within each region  ${\cal R}_\pm$.  In the RW gauge, these equations can be decoupled in terms of two complex master functions (the RW master function for axial perturbations and the ZM master function for polar perturbations) that satisfy wave-type master equations with the structure of Eq.~(\ref{eq:3.1}) but without source terms. Then, we have to solve simultaneously the equations for $\Psi^{\RW/\ZM,+}_{\ell m}$ (for $r>r_{p}$) and for $\Psi^{\RW/\ZM,-}_{\ell m}$ (for $r<r_{p}$) with the jump conditions, $\salto{\Psi^{\RW/\ZM}_{\ell m}}$ and $\salto{\partial^{}_{r}\Psi^{\RW/\ZM}_{\ell m}}$, imposed at $r=r_{p}$. These jumps are just time-dependent functions of the SCO dynamics (including its energy-momentum tensor), whose expression will be given in~\cite{Canizares:2014cs} (but can be derived easily from already existing works, e.g.~\cite{Sopuerta:2005gz}).  Finally, we reconstruct all the metric perturbations in the RW gauge at both sides of the SCO trajectory following the usual procedure and using the jump conditions imposed by the perturbative Einstein equations.

-\textit{Computing the GSF.} The GSF is well-defined in the L gauge after the developments of~\cite{Mino:1997nk,Quinn:1997am}.  Its expression in terms of the (reversed) metric perturbations is:
\begin{eqnarray}
F^{\mu}_{\LO} = -\frac{m_{\ast}}{2}\left( {\met}^{\mu\nu}_{\BH} + u^{\mu}u^{\nu}\right)\left( 2\,\psi^{\LO,\REG}_{\nu\rho;\sigma} - \psi^{\LO,\REG}_{\rho\sigma;\nu} \right)u^{\rho}u^{\sigma}\,,
\label{eq:5.1}
\end{eqnarray}
where the superscript R refers to the fact that the metric perturbations used here must have been regularized to substract the singular field that does not contribute to the SCO motion.

In order to compute the GSF in the L gauge, we need to find a gauge vector $\xi^{\mu}$ that takes us from the RW gauge to the L gauge. Then, the only thing left is to find a formulation where we can decouple the equations for the components of $\xi^{\mu}$. From Eq.~(\ref{gaugetransformationequation}) and the L gauge condition we get the following equations: $\xi^{}_{\mu;\nu}{}^{;\nu} = \psi^{\RW}_{\mu\nu}{}^{;\nu}\,$. The source terms of this set of inhomogeneous wave-type equations are made out of RW metric perturbations.   We have not been able to decouple this set of equations.  However, following the computation of the electromagnetic self-force on eccentric geodesics in Schwarzschild presented in~\cite{Haas:2011np}, where the equations for the electromagnetic vector $A^{\mu}$ are decoupled, we can find a way of decoupling the equations for $\xi^{\mu}$.  The starting point of~\cite{Haas:2011np} are Maxwell's equations: $F^{\mu\nu}{}^{}_{;\nu} = 4\pi j^{\mu}_{\rm EM}$ with $F^{}_{\mu\nu} = A^{}_{\nu,\mu}-A^{}_{\mu,\nu}$.   In order to do the same for $\xi^{\mu}$ we need to specialize to a particular L gauge, such that when we construct equations of the Maxwell type, i.e. $\left(\xi^{}_{\nu,\mu}-\xi^{}_{\mu,\nu}\right)^{;\nu}= {\cal S}^{}_{\mu}$, the source term ${\cal S}^{}_{\mu}$ only contains components of the RW metric perturbations.  It turns out that ${\cal S}^{}_{\mu}$ contains the trace of the metric perturbations in the L gauge, $h^{\rm L} = g^{\mu\nu}h^{\rm L}_{\mu\nu}$, which we do not know a priori.   Actually, this quantity, outside matter sources satifies: $h^{\rm L}{}_{;\mu}{}^{;\mu}=0\,$.  It is well-known that in vacuum we can specialize to a L gauge where the trace-free condition $h^{\rm L}=0$ holds.  Since working with the PwP formulation allows us to work with the vacuum Einstein equations at both sides of the SCO trajectory, we can adjust the gauge to have $h^{\rm L,\pm}=0$. Then, we can use the following equations for $\xi^{\mu}_{\pm}$:
\begin{eqnarray}
\xi^{\pm}_{\mu}{}^{}_{;\nu}{}^{;\nu} - \xi^{\pm}_{\nu}{}^{;\nu}{}_{;\mu}= h^{{\RW},\pm\;}_{\mu\nu}{}^{;\nu} - h^{{\RW},\pm}_{;\mu}\,.
\label{equationsforxi}
\end{eqnarray}
The equations for the polar sector can be decoupled by introducing the following combinations of the components of $\xi^{\pm}_{\mu}$: $\chi^{\pm}_{1} = -r^{2}(\xi^{\pm}_{t,r}-\xi^{\pm}_{r,t})\,$, $\chi^{\pm}_{2} = f(\xi^{\pm}_{r}-\xi^{\pm}_{{\cal P},r})\,$, and $\chi^{\pm}_{3} = \xi^{\pm}_{t}-\xi^{\pm}_{{\cal P},t}\,$.  The master equations they satisfy are:
\begin{eqnarray}
\left(\Box^{}_{2}-V^{}_{\ell}(r)\right)\chi^{\pm}_{1,2} = {\cal S}^{\RW,\pm}_{1,2}[h^{\RW,\pm}_{\mu\nu}]\,,
\label{masterxi12}
\end{eqnarray}
\begin{eqnarray}
\left(\Box^{}_{2}-V^{}_{\ell}(r) \right)\chi^{\pm}_{3} -V^{13}_{\ell}(r)\chi^{\pm}_{1}= {\cal S}^{\RW,\pm}_{3}[h^{\RW,\pm}_{\mu\nu}]\,, 
\label{masterxi3}
\end{eqnarray}
where $V^{}_{\ell}(r)=-\ell(\ell+1)f/r^2$ and $V^{13}_{\ell}(r)= f(3f-1)/r^3$.  The only axial component satisfies an equation like~(\ref{masterxi12}) with a source term ${\cal S}^{\RW,\pm}_{4}[h^{\RW,\pm}_{\mu\nu}]$ (the expressions of these sources will be given in~\cite{Canizares:2014cs}).  These equations again have to be solved simultaneously at each side of the SCO trajectory and the solutions have to be connected through the jumps conditions on $\chi^{\pm}_{I}$ ($I=1-4$).  These jumps, $\salto{\chi_{I}}$, can be derived from the ones in $\xi^{\mu}$, and will be given in~\cite{Canizares:2014cs}.  Notice that in order to solve Eq.~(\ref{masterxi3}) for $\chi^{\pm}_{3}$ we first need to solve Eq.~(\ref{masterxi12}) for $\chi^{\pm}_{1}$. Once the equations are solve, we can construct the components of $\xi^{\mu}$ and then the metric perturbations in the L gauge.  This finishes the program for computing the GSF based on computations in the RW gauge that only involve solving individual wave equations.

-\textit{Discussion and conclusions}. We can try to make the procedure described here even more efficient if we would have a prescription for computing the GSF in the RW gauge.  As we have seen, although the gauge vector to transform from the L to the RW gauge is smooth at both sides of the SCO trajectory and finite, the values on the trajectory itself are different coming from one side or the other.  These jumps in the gauge vector generate singularities in the RW metric perturbations in the form of Dirac delta distributions.  Although these terms are crucial to deal with the singularities in the field equations for the RW metric perturbations, they are too singular to use any of the presently known formulations of the self-force.  Perhaps one can use methods based on averaging like ones in~\cite{Pound:2013faa,*Gralla:2011ke,*Gralla:2011zr}.  In any case this is a question that deserves further investigation.

On the other hand, the techniques described in this letter may be of great help for the case of Kerr.  We can use similar techniques in order to  try to compute the metric perturbations in gauges where the computational cost is comparable with solving master equations plus a reconstruction procedure. In particular, one can consider radiation-type gauges.  Regarding actual computations, it is important to remark that the PwP has already been implemented in practice in the time domain pseudospectral collocation method~\cite{Canizares:2008dp,Canizares:2009ay,Canizares:2010yx,Canizares:2011kw,Jaramillo:2011gu}.  Implementations in the frequency domain are also possible.  In any case, the fact that the PwP allows us to work with homogeneous equations, avoiding source terms with low differentiability, ensures the spectral numerical convergence of this type of numerical implementation.

\begin{acknowledgements}
We thank the members of the CAPRA community for enlightening discussions over the years.  PC is supported by FP7 Marie Curie Intra-European contract PIEF-GA-2011-299190.  CFS is supported by contracts AYA-2010-15709 (Spanish Ministry of Science and Innovation) and 2009-SGR-935 (AGAUR, Generalitat de Catalunya).
\end{acknowledgements}


%

\end{document}